\documentclass[journal, twoside, web]{ieeecolor}

\usepackage{geometry}
\usepackage{amsmath}
\usepackage{amsfonts}
\usepackage{graphicx}
\usepackage{ amssymb }
\usepackage{bbold}
\let\labelindent\relax
\usepackage{enumitem}
\usepackage{titlesec}
\usepackage{xcolor}

\newtheorem{prop}{Proposition}
\newtheorem{defn}{Definition}

\newtheorem{rem}{Remark}
\newtheorem{exmp}{Example}
\newtheorem{thm}{Theorem}

\newtheorem{assumption}{Assumption}

\def\-{\raisebox{.0pt}{-}}

\newcommand{\R}{\mathbb{R}}

\newcommand{\B}[1]{\boldsymbol{#1}}

\newcommand{\be}{\begin{equation}}
\newcommand{\ee}{\end{equation}}
\newcommand{\ba}{\begin{aligned}}
\newcommand{\ea}{\end{aligned}}

\newcommand{\lba}{\left[ \begin{array}}
\newcommand{\ear}{\end{array} \right]}

\newcommand{\bA}{\overline{A}}
\newcommand{\bB}{\overline{B}}

\newcommand{\sm}{\small{-}}

\title{Certifying Stability and Performance of Uncertain Differential-Algebraic Systems: A Dissipativity Framework}
\author{Emily Jensen, \IEEEmembership{Member, IEEE}, Neelay Junnarkar, \IEEEmembership{Student Member, IEEE}, Murat Arcak, \IEEEmembership{Fellow, IEEE}, Xiaofan Wu, Suat Gumussoy, \IEEEmembership{Senior Members, IEEE}
\thanks{This work was supported in part by {Siemens Corporation R\&D funding} and by the NSF grant CNS-2135791.}
\thanks{Emily Jensen, Neelay Junnarkar, and Murat Arcak are with the department of Electrical Engineering and Computer Sciences at the University of California, Berkeley. (emails: $\{$emilyjensen, neelay.junnarkar, arcak$\}$ @berkeley.edu)}
\thanks{Xiaofan Wu and Suat Gumussoy are with Siemens Technology. (emails: $\{$xiaofan.wu, suat.gumussoy$\}$ @siemens.com}}

\begin{document}
\maketitle

\begin{abstract}
This paper presents a novel framework for characterizing dissipativity of uncertain systems whose dynamics evolve according to differential-algebraic equations.
Sufficient conditions for dissipativity (specializing to, e.g., stability or $L_2$ gain bounds) are provided in the case that 
uncertainties 
are
characterized by integral quadratic constraints. For polynomial or linear dynamics, these conditions can be efficiently verified through sum-of-squares or semidefinite programming. Performance analysis of the IEEE 39-bus power network with a set of potential line failures modeled as an uncertainty set provides an illustrative example that highlights the computational tractability of this approach; conservatism introduced in this example is shown to be quite minimal.
\end{abstract}

\vspace{-5mm}

\section{Introduction}

Dissipativity theory \cite{willems1972dissipative} relates input-output properties of dynamical systems to the dissipation of 
so-called {\it storage functions} over  trajectories
\cite{lozano2013dissipative}. 
Storage functions generalize Lyapunov functions to systems with inputs and outputs:
 the time derivative of the storage function along trajectories is upper bounded by a \emph{supply rate} that describes a relation of the system's inputs and outputs. 
{Appropriate choices of supply rate lead to important cases of dissipativity, such as passivity and stability, as well as $L_2$ gain bounds ($H_{\infty}$ norm) which is key to robustness analysis.} 
Moreover, 
a dissipativity framework 
 allows for incorporation of uncertainties described by integral quadratic constraints (IQCs) \cite{hu2017robustness, megretski1997system}. {This will be key to developing our main result, which provides a sufficient condition for dissipativity of uncertain systems described by differential algebraic equations (DAEs).}

Classical dissipativity applies to system dynamics described by ordinary differential equations (ODEs), {and sufficient conditions for dissipativity of uncertain ODE systems have been developed \cite{xie1998robust}.}  Despite the vast literature on dissipativity and on DAEs,  a {general dissipativity framework} for {uncertain} and nonlinear differential-algebraic systems is lacking, and this work addresses this gap. 

DAEs model dynamical systems
with {inherent algebraic} constraints and arise in chemical,
electrical, and mechanical engineering applications \cite{kumar1999control},  
e.g., {the inherent conservation of charge} can be modeled as a constraint in electric circuit dynamics, 
{imcompressibility can be captured as a constraint in} fluid flow modeled by the Navier-Stokes equations, 
or { holonomic constraints can be used to model }mechanical system motion 
\cite{ascher1998computer}.
Many classical results have been extended from the ODE 
to the DAE setting, including controllability and observability \cite{yip1981solvability}, and  controller \cite{feng2017robust} and observer \cite{darouach1995design} design for linear systems. Specific subclasses of dissipativity that have been examined for DAEs include positive realness of linear DAE 
systems \cite{chu2008algebraic} and Lyapunov stability and passivity of nonlinear DAE 
systems \cite{liu2008passivity}. 
However, incorporation of uncertainties into dissipativity analysis of DAE systems is less prevalent, {and this work addresses this gap.} One exception is \cite{uezato1999strict} which analyzes stability of linear DAE systems subject to polytopic uncertainties.

This work is motivated by DAE models with algebraic constraints that model underlying {network 
interconnections}, e.g., dynamics of multibody systems with interconnection constraints \cite{udwadia2006explicit}, power networks with algebraic constraints describing the power flow equations \cite{7585069}, and implicit neural networks with outputs defined as the solutions to fixed-point equations \cite{ghaoui2020implicit}. When the algebraic constraint is invertible, the DAE system may be equivalently represented as a standard ODE, e.g., power network DAEs can be converted to an ODE through a Kron reduction procedure \cite{dorfler2012kron}. {However, when the algebraic constraint  captures underlying graph properties, it may be desirable to preserve this form to incorporate uncertainties or changes to the graph in a way that preserves the graph structure. In other cases, the algebraic conditions may not be invertible or an inversion may be ill-conditioned or computationally challenging.}

In this manuscript, we provide \emph{sufficient conditions for dissipativity of DAE systems with uncertainties described by
quadratic constraints.}
For linear or polynomial dynamics, we show that \emph{these conditions can be verified numerically with linear matrix inequalities or Sum-of-Squares programming,} respectively. {Throughout, we emphasize computational tractability of the conditions derived, at the expense of some conservatism. We illustrate the usefulness of this choice in Section~\ref{sec:case_study}, where the performance of a power network subject to an uncertain line failure is quickly verified.} 

The remainder of the paper is structured as follows. In Section~\ref{sec:setup}, the DAE model of interest with uncertainties characterized by IQCs is presented; the notion of dissipativity for this model is formalized. 
Section~\ref{sec:general} presents a sufficient condition for dissipativity of uncertain DAE systems. 
Numerical methods to confirm this condition are noted in 
Section~\ref{sec:SOS} and Section~\ref{sec:linear} for polynomial and linear dynamics, respectively.
A power network with an uncertain line failure is analyzed as a case study in Section~\ref{sec:case_study}.

\vspace{-5mm}

\section{Problem Set-Up: Uncertain Differential-Algebraic Systems} \label{sec:setup} 
{We consider a time-invariant
system 
with 
state dynamics described by an ordinary-differential equation along with an algebraic condition:}
    \begin{subequations} \label{eq:nonlin_dae}
    \begin{align} 
        \dot{\boldsymbol{ x}}(t) &= f(\boldsymbol{x}(t), \boldsymbol{ v}(t), \boldsymbol{ w} (t), \boldsymbol{ \xi}(t)) \label{eq:nonlin_ode}\\
        0 & = g( \boldsymbol{x}(t),\boldsymbol{v}(t), \boldsymbol{w}(t), \boldsymbol{\xi} (t)) \label{eq:nonlin_const}\\
        \boldsymbol{y}(t) &= h(\boldsymbol{x}(t),\boldsymbol{v}(t)),
    \label{eq:nonlin_output} 
    \end{align}
    \end{subequations}
where
$\B{x}(t) \in \R^n$ is the state, $\B{v}(t) \in \R^m$ is the algebraic variable, $\B{w}(t) \in \R^p$ is an exogenous disturbance {with finite energy ($\|w\|_{L_2} < \infty$)}, and $\B{y}(t) \in \R^q$
is an output. 
$\B{\xi}(t) \in \R^{\ell}$ captures additional dynamics, e.g., uncertainties, and is modeled as the output of a bounded, causal system $\Delta:L_{2}^n \times L_{2} ^m\rightarrow L_{2}^{\ell}$:
\vspace{-2mm}
    \be\label{eq:Delta}
        \B{\xi} = \Delta \left( \B{x}, \B{v}\right).\vspace{-1mm}
    \ee
 Bold letters denote signals, ${\boldsymbol{x}}: [0, \tau] \rightarrow \R^n$,  and non-bold letters denote points, $x \in \R^n$. 
We assume an equilibrium point exists and is shifted to $x=0$; i.e., 
$f(0,v_0,0,0) = 0, g(0,v_0,0,0) = 0$ for some $v_0.$
For $\B{\xi} = 0$, \eqref{eq:nonlin_dae} is a 
\emph{differential-algebraic equation} (DAE) in \emph{semi-explicit form} \cite{ascher1998computer,kumar1999control}.

{
\begin{assumption} \label{assumption}
The initial conditions of \eqref{eq:nonlin_dae} are \emph{consistent}, e.g., they satisfy the {algebraic condition} \eqref{eq:nonlin_const}.
Inputs to the system are sufficiently smooth\footnote{When $\frac{\partial g}{\partial v}$ is nonsingular, the DAE is of \emph{index 1} \cite{ascher1998computer}; for higher index systems, the solution of \eqref{eq:nonlin_dae} will be dependent on derivatives of the input $w$ \cite{kumar1999control}.}. 
\end{assumption}
}

Under Assumption 1, a unique solution $\B{x}, \B{v}, \B{\xi}, \B{y}$, exists for system \eqref{eq:nonlin_dae}-\eqref{eq:Delta} over an interval of time $t \in [0,\tau]$. 
Methods for determining admissible initial conditions can be found in, e.g., \cite{lin1988existence}, and existence and uniqueness of solutions can be confirmed through, e.g., geometric approaches \cite{reich1991existence}, \cite{rabier1994geometric}, theory of differential equations on manifolds \cite{rheinboldt1984differential}, or computational methods \cite{campbell1995solvability}. Further details are beyond the scope of this work; we refer the reader to \cite{ascher1998computer, kumar1999control} and the references therein for a more comprehensive study.

\subsection{Dissipativity of DAE Systems}
We begin by formalizing the notion of dissipativity for uncertain DAE systems. 

\begin{defn}\label{defn:dissipativity}
Under Assumption~\ref{assumption}, the DAE system \eqref{eq:nonlin_dae}-\eqref{eq:Delta} is \emph{dissipative} with respect to the supply rate $s(\cdot, \cdot)$ if there exists a \emph{storage function} $V$, that is positive and satisfies $V(0) = 0$ and
    \be 
        \label{eq:dissipative_integral}
            V(\B{x}(T)) - V(\B{x}(0)) \le \int_0^T s(\B{w}(t), h(\B{x}(t), \B{v}(t))) dt
    \ee 
for all $T \in [0, \tau].$ 

\end{defn}

Note that $\B{x}, \B{w}, \B{v},$ and $\B{\xi}$ in \eqref{eq:dissipative_integral} satisfy the dynamics \eqref{eq:nonlin_dae}.
Supply rates of interest include: 
    \begin{itemize}
        \item $s(w,y) = w^{\top}y$ corresponding to \emph{passivity},
        \item $s(w,y) = w^{\top} w - \gamma^2y ^{\top} y$ corresponding to an \emph{$L_2$ gain bound} of $\gamma$,
        \item $s(w,y)\equiv0$ corresponding to \emph{stability} of the origin; in this case, the storage function $V(\cdot)$ serves as a Lyapunov function.
    \end{itemize}

{As with the standard ODE setting,} this definition is difficult to verify {in uncertain settings,} i.e., with presence of unknown $\Delta$. In what follows, we derive sufficient conditions for dissipativity when $\Delta$ is unknown, but satisfies known quadratic constraints.

\vspace{-3mm}
\subsection{Integral Quadratic Constraints}
\begin{figure}[t]
    \centering
    \includegraphics[width = .4\textwidth]{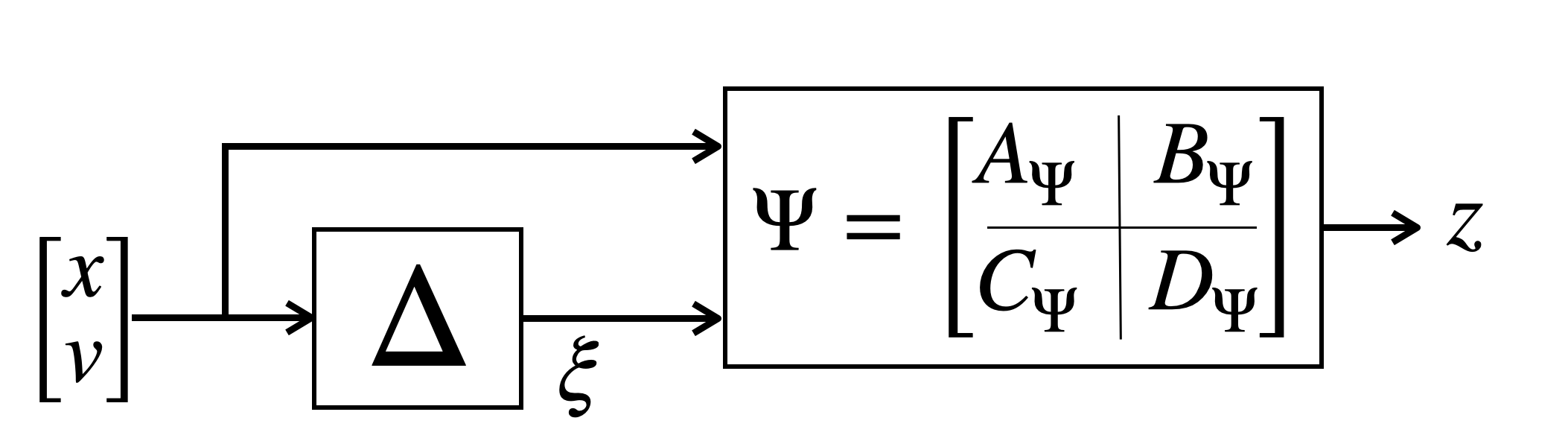}\vspace{-3mm}
    \caption{Block diagram representation of uncertain system $\Delta$ and virtual filter $\Psi$. $\Psi$ is  utilized to provide a more general input-output characterization of $\Delta$ via the integral quadratic constraint $\int_0^{\infty} z(t)^{\top}M z(t) dt \le 0$ for some $M = M^{\top},$ where $z$ is the output of $\Psi$. } \vspace{-5mm}
    \label{fig:iqc}
\end{figure}

{Incorporation of uncertainties is a key component of our framework, which will develop sufficient conditions for dissipativity of \emph{uncertain} DAE systems. Toward this aim, we utilize an integral quadratic constraint framework, which has previously been incorporated into dissipativity analysis in the ODE setting \cite{hu2017robustness}.}

We characterize $\Delta$ in \eqref{eq:Delta} through input-output properties with the framework of integral quadratic constraints (IQCs) \cite{megretski1997system, seiler2010dissipation}. As depicted in Figure~\ref{fig:iqc}, the input signals $\B{x},\B{v}$ and output signal $\B{\xi}$ of $\Delta$ are passed through a ``virtual filter" $\Psi$ defined by the stable linear dynamics:
    \be \ba \label{eq:filter_dynamics}
        \dot{\B{\psi}}(t) &= A_{\Psi} \B{\psi}(t) + B_{\psi} {\small{\lba{c} \B{x} ( t) \\ \B{v}(t) \\ \B{\xi}(t) \ear}}, ~~ \B{\psi}(0) = 0 \\
        \B{z}(t) & = C_{\Psi} \B{\psi}(t) + D_{\Psi}{\small{\lba{c} \B{x} ( t) \\ \B{v}(t) \\ \B{\xi}(t) \ear}}.
    \ea \ee
For $M = M^{\top}$, $\Delta$ \emph{satisfies the hard IQC defined by $(\Psi, M)$} if
    \be \label{eq:general_iqc}
        \int_0^{T} \B{z}(t)^T M \B{z}(t) dt \le 0 
    \ee 
for all $T \ge 0$. This is clearly satisfied if 
    \be \label{eq:pointwise_static_IQC}
        \B{z}(t)^{\top}  M \B{z}(t) \le 0, ~~ \forall t \ge 0,
    \ee 
and we say $\Delta$ satisfies the \emph{pointwise quadratic constraint} defined by $(\Psi,M)$.
In the simple case that $\Psi$ is the identity operator, 
\eqref{eq:general_iqc} reduces to
    \be \label{eq:static_IQC}
        \int_0^{T} {\small{\lba{c} \B{x}(t) \\ \B{v}(t)\\ \B{\xi}(t) \ear}}^{\top} M {\small{\lba{c} \B{x}(t) \\ \B{v}(t)\\ \B{\xi}(t) \ear}} dt \le 0.     \ee 

\vspace{-2mm}

\section{Conditions for Dissipativity of Differential-Algebraic Systems} \label{sec:general} 
\vspace{-1mm}
This section presents a sufficient condition for dissipativity of DAE systems with uncertainties described by IQCs; a method for confirming this condition numerically in the case of polynomial DAEs is illustrated.

\begin{thm} \label{thm:nonlin_dissipativity}
Consider the DAE system \eqref{eq:nonlin_dae}-\eqref{eq:Delta} and assume $\Delta$ satisfies the IQC defined by $(\Psi,M)$. 
This system is dissipative w.r.t. the supply rate $s(\cdot, \cdot)$ if there exist $\tau,\lambda \ge 0$, a matrix $P_{\Delta} \succeq 0$, and a positive definite $V(\cdot)$ satisfying $V(0) = 0 $ and 
    \be \ba \label{eq:nonlinear_general_int}
       & \triangledown V(x)^{\top} f(x,v,w,\xi)+  {\psi}^{\top} P_{\Delta} \Big( A_{\psi} \psi\small{+} B_{\psi} \lba{c}x \\ {v} \\ {\xi} \ear  \Big) \\
       &+ \Big( A_{\psi} \psi\small{+} B_{\psi} \lba{c}x \\ {v} \\ {\xi} \ear  \Big) ^{\top} P_{\Delta} \psi \\
       & \le~ s\big(w, h(x,v)\big)+\lambda  g(x,v,w,\xi) ^{\top}g(x,v,w,\xi) ~+ \\
       &\tau\Big( C_{\psi} \psi + D_{\psi} {\small{\lba{c} x\\ v \\ \xi \ear}} \Big)^{\top} M \Big( C_{\psi} \psi + D_{\psi} {\small{\lba{c} x\\ v \\ \xi \ear}} \Big),
    \ea \ee 
for all $\psi,x,v,\xi,w.$
\end{thm}

{The left hand side of inequality \eqref{eq:nonlinear_general_int} is the derivative of a storage function \be \label{eq:combined_storage}
        \tilde{V}(x,\psi): = V(x) + \psi^{\top}P_{\Delta} \psi
    \ee 
for the augmented system of DAE \eqref{eq:nonlin_dae} and filter \eqref{eq:filter_dynamics}. The terms on the right hand side of the inequality following the supply rate arise from the algebraic constraint \eqref{eq:nonlin_const} and the incorporation of uncertainty.} 
{When uncertainty is characterized by a \emph{pointwise} constraint, we may assume $P_{\Delta} = 0.$} Nonzero $P_{\Delta}$ 
introduces an additional 
term to the left hand side of \eqref{eq:nonlinear_general_int} whose negativity may help
this inequality hold.

{
\textit{Proof of Theorem~\ref{thm:nonlin_dissipativity}:} 
Integrating \eqref{eq:nonlinear_general_int} from $t = 0$ to $T$ along trajectories of \eqref{eq:nonlin_dae} and \eqref{eq:filter_dynamics} and using the filter initial condition $\B{\psi}(0)=0$ gives
    \be \ba \label{eq:pf1}
       &  V(\B{x}(T)) + \B{\psi}(T)^T P_{\Delta} {\B{\psi}}(t) - V(\B{x}(0)) \\
       & - \int_0^T s\big(\B{w}(t), h\big(\B{x}(t),\B{v}(t)\big)\big) dt  \\
       &\le ~  \lambda \int_0^T (\star)^{\top}g(\B{x}(t),\B{v}(t),\B{w}(t),\B{\xi}(t)) dt ~+\\
       & ~~  \tau \int_0^T (\star)^{\top} M \Big( C_{\psi} \B{\psi}(t) + D_{\psi} {\small{\lba{c} \B{x}(t) \\ \B{v}(t) \\ \B{\xi}(t) \ear}} \Big)dt,
    \ea \ee 
where terms denoted by $(\star)$ can be inferred by symmetry. 
Non-positivity of the terms on the right hand side of \eqref{eq:pf1} follow from \eqref{eq:nonlin_const} and that $\Delta$ satisfies the IQC defined by $(\Psi, M)$. 
By the classical S-procedure, non-positivity of these terms imply
\begin{equation} \ba \label{eq:1} & V(\B{x}(T)) - V(\B{x}(0)) + \B{\psi}(T)^T P_{\Delta} {\B{\psi}}(T) ~ \le   \\
      &  \int_0^T s\big(\B{w}(t), h\big(\B{x}(t),\B{v}(t)\big)\big) dt.  \ea \end{equation}
\eqref{eq:dissipative_integral} follows from \eqref{eq:1} since $P_{\Delta} \succeq 0$. \hfill $\blacksquare$
}

{A source of conservatism for Theorem~\ref{thm:nonlin_dissipativity} arises from requiring 
a \emph{single} storage function
$\tilde{V}(\cdot, \cdot)$  to ensure dissipativity \emph{for all} $\Delta$ satisfying a quadratic constraint. This conservatism allows us to avoid the computational burden associated with finding a separate storage function for each separate $\Delta$ satisfying the IQC. Additional conservatism arises because the S-procedure applied is, in general, not lossless.}

Note that the condition \eqref{eq:nonlinear_general_int} for dissipativity simplifies in the case of no uncertainty to:
        \be  \ba\label{eq:dae_dissipative_integral}
       &  \triangledown V(x)^{\top} f(x,v,w,0) \le\\
       &s\big(w, h(x,v)\big) + \lambda  g(x,v,w,0)^{\top} g(x,v,w,0).
      \ea   \ee 

{Theorem~\ref{thm:nonlin_dissipativity} is related to the stability analysis of linear DAE systems with polytopic uncertainties presented in \cite{uezato1999strict}. Theorem~\ref{thm:nonlin_dissipativity} extends this notion to more general notions of dissipativity and allows for nonlinear dynamics.} 

 \vspace{-3mm}
\subsection{Numerical Solutions for Polynomial Dynamics}
\label{sec:SOS}

When the DAE system is described by polynomials, the storage function \eqref{eq:combined_storage} can be found numerically with a sum-of-squares approach.

A polynomial $p$ is \emph{sum-of-squares} (SOS) if there exist polynomials $p_1, \dots, p_n$ such that $p = \sum_{i=1}^n p_i^2$. 
A polynomial being SOS implies it is nonnegative. Let $\Sigma[x]$ be the set of SOS polynomials in $x$, and $\Sigma[(x, v, w, \xi, \psi)]$ be the set of SOS polynomials in $x, v, w, \xi$ and $\psi$. Then, for polynomial $f, g,$ and $h$ and a polynomial supply rate $s$, we can verify dissipativity through Theorem~\ref{thm:nonlin_dissipativity} by finding nonnegative scalars $\tau$ and $\lambda$, a positive definite matrix $P_\Delta$, and a $V \in \Sigma[x]$ such that $V(x) - \epsilon x^\top x  \in \Sigma[x]$ and
\be
\ba \label{eq:SOS_general}
    s(w, h(x, v)) + \lambda \cdot (\star)^\top g(x, v, w, \xi) \\ + \tau \cdot (\star)^\top M \Big(C_\psi \psi + D_\psi  {\small{\lba{c} x\\ v \\ \xi \ear}}\Big) \\ - \triangledown V(x)^\top f(x, v, w, \xi) - \psi^\top P_\Delta \psi \\ - \psi^\top P_\Delta \Big(A_\psi \psi + B_\psi  {\small{\lba{c} x\\ v \\ \xi \ear}}\Big) \\ \in \Sigma[(x, v, w, \xi, \psi)]\\
\ea
\ee
where \(\epsilon\) is small and positive and each of the terms $(\star)$ can be inferred by symmetry.

\begin{exmp}
Consider the polynomial DAE system:
\be \ba  \label{eq:exmpSOS}
    \dot{\B{x}}_1(t) &= -\B{x}_1(t) + \B{v}(t)  \\ 
    \dot{\B{x}}_2(t) &= -\B{x}_1(t) -\B{x}_2(t) \\
    0 & = \B{x}_1(t)^2 + \left(\B{x}_2(t)^2 + 5 \right)\B{v}(t). \\
\ea \ee 
Applying \eqref{eq:SOS_general} to this system to show stability of the origin gives the following equations
\begin{equation*}
\ba 
    V(x) - \epsilon x^\top x & \in \Sigma[x], \\
    \lambda g(x, v)^\top g(x, v)  
    - \triangledown V(x)^\top f(x, v) & \in \Sigma[(x, v)].
\ea
\end{equation*}
To implement this, we use the SOSTOOLS MATLAB toolbox \cite{prajna2002introducing} and the SeDuMi solver \cite{sturm1999using}. We allow $V$ to be a polynomial of degree $\le4$. For $\epsilon = 10^{-3}$, SeDuMi finds a solution $\lambda = 0.59504$ and $V(x) = 0.00017634x_1^4 + 0.0012261x_1^2 x_2^2 
    + 0.0027498x_1 x_2^3 + 0.0023039x_2^4 + 0.013246x_1^3
    - 0.013733x_1^2x_2 - 0.055089x_1x_2^2 - 0.056305x_2^3
    + 0.40316x_1^2 + 0.67688x_1x_2 + 0.57717x_2^2$.
\end{exmp}

\vspace{-2mm}
\section{Dissipativity of Linear Differential-Algebraic Systems} \label{sec:linear}
For linear DAE dynamics, dissipativity can be verified 
with feasibility of an LMI.
The linear DAE model is: 
    \begin{subequations}\label{eq:lin_uncertain}
        \begin{align} 
            &\dot{\B{x}}(t) {\small{=}}  A \B{x}(t) {\small{+}} B_v \B{v}(t) + B_w \B{w}(t) + B_{\xi} \B{\xi}(t) \label{eq:lin_ODE}\\
            &0  { =} F \B{x}(t) + G_v\B{v}(t) + G_w \B{w}(t) + G_{\xi} \B{\xi}(t) \label{eq:lin_const}\\
            &\B{y}(t)  = C \B{x}(t) + D_v \B{v}(t),
    \end{align} \end{subequations}
    \vspace{-1mm}
where \vspace{-1mm}
    \be \label{eq:Delta2}
        \B{\xi}= \Delta(\B{x},\B{v})
    \ee 
can describe nonlinearities such as model uncertainties \cite{buch2021robust} or saturations \cite{hindi1998analysis}. We restrict our attention 
to \emph{quadratic} supply rates: \vspace{-1mm}
    \begin{equation*}
            s(w,y)  = \lba{c} y \\ w\ear^{\top} \tilde{X}_s \lba{c} y \\ w\ear\\
    \end{equation*}
which, using the equality $y = Cx + D_v v$, can be written as \vspace{-1mm}
\setlength{\arraycolsep}{1.5pt}
    \be \ba \label{eq:quad_s}
              s(w,Cx \small{+} D_v v) {\small{=}} \lba{c} x \\ w \ear^{\top} \lba{cc} X_{xx} & X_{xw} \\ X_{xw}^{\top} & X_{ww} \ear  \lba{c} x \\ w  \ear
        \ea \ee 
For linear dynamics and quadratic supply rates, there is no loss \cite{willems1972dissipative} in restricting a storage function $V(\cdot)$ to be quadratic: 
    \be  \label{eq:quadratic_V}
        V(x) = x^{\top} P x, ~~ P = P^{\top} \succ 0.
    \ee 
\vspace{-5mm}

The proof of the following result is presented in Appendix B.

\begin{prop} \label{thm:linear_general}
Consider the linear DAE system \eqref{eq:lin_uncertain}-\eqref{eq:Delta2} with $ \Delta$ satisfying the IQC defined by $(\Psi,M)$. 
This system is dissipative w.r.t. the quadratic supply rate \eqref{eq:quad_s} if there exists $\lambda, \tau \ge0$, $P \succ0$, and $P_{\Delta} \succeq 0$ 
satisfying
\setlength{\arraycolsep}{3pt}
\be\ba \label{linear_thm_eq}
    & \lba{cc|c} 
        X(P) & P B_w  & B_{\psi}^{\top} P_{\Delta}\\
         B_w^{\top} P & 0  &  0 \\ \hline P_{\Delta} B_{\psi} & 0 &  A_{\psi}^{\top} P_{\Delta} + P_{\Delta} A_{\psi} 
    \ear 
     \\
    & ~~~~~\preceq\lambda (\star)^{\top} \lba{cccc|c} F & G_v & G_{\xi} & G_w & 0 \ear \\
    & ~~~~~~~~+ \tau \lba{cc|c} D_{\psi}^{\top} M D_{\psi} & 0 & D_{\psi}^{\top} M C_{\psi} \\ 0 &0 &0 \\ \hline C_{\psi}^{\top}M D_{\psi} & 0 & C_{\psi}^{\top} M C_{\psi} \ear,
\ea \ee 
where {\small{$X(P): = \lba{ccc} A^{\top}P + P A & P B_v & P B_{\xi} \\ B_v^{\top} P & 0 & 0 \\ B_{\xi}^{\top}P & 0& 0 \ear $}.} (The matrices in \eqref{linear_thm_eq} are block partitioned to assist with readability.)
\end{prop}

\begin{rem} Conservatism of Proposition~\ref{thm:linear_general} arises from requiring 
a \emph{single} storage function
$\tilde{V}(\cdot, \cdot)$ to ensure dissipativity \emph{for all} $\Delta$ satisfying a quadratic constraint. {The benefit of this is the computational efficiency of considering feasibility of a 
single LMI to
confirm dissipativity over the full uncertainty set.}
\end{rem}

When the uncertainties can be captured as the output of a system $\Delta$ satisfying a \emph{pointwise} quadratic constraint defined by $(I,M)$, the parameter $P_{\Delta}$ accounting for the filter dynamics
can be taken as zero, and condition \eqref{linear_thm_eq} simplifies to 
    \setlength{\arraycolsep}{3pt}
        \be \ba \label{eq:lin_uncert_LMI}
            & \lba{cccc} A^{\top} P + P A & P B_v & P B_{\xi} & P B_w \\ 
            B_v^{\top} P & 0 & 0 & 0\\
            B_{\xi}^{\top} P & 0 & 0 & 0\\
            B_{w}^{\top} P & 0 & 0 & 0 
            \ear \preceq \tau \tilde{M} ~+\\ 
           & ~~~~~~~ \lba{cccc} X_{xx} & 0 & 0 & X_{xw} \\
           0 & 0 & 0 & 0 \\
           0 & 0 & 0 & 0\\
            X_{xw}^{\top} &  0& 0  &X_{ww}\ear+ \lambda \lba{c} F^{\top} \\ G_v^{\top} \\ G_{\xi}^{\top} \\ G_{w}^{\top} \ear \lba{c} F^{\top} \\ G_v^{\top} \\ G_{\xi}^{\top} \\ G_{w}^{\top} \ear ^{\top}
        \ea \ee 
where $\tilde{M} := \lba{cc} M & 0 \\ 0 & 0\ear.$

\begin{exmp} \label{exmp:nn}
{\bf (Implicit neural network controller)}

Consider a general linear plant with dynamics 
    \be \ba \label{eq:nn_plant}
        \dot{\B{x}}_p(t) &= A_p \B{x}_p(t) + B_u \B{u}(t) +B_w \B{w}(t)\\
        \B{y}(t) & = C_p \B{x}_p(t)
    \ea \ee 
in feedback with a controller $\pi_{\theta}$ modeled as the interconnection of an LTI system 
and activation functions $\phi$: 
    \be \ba
        \dot{\B{x}}_k(t) &= A_k \B{x}_k(t) + B_{\xi} \B{\xi}(t) +  B_{y} \B{y}(t) \\
        \B{u}(t) &= C_{u} \B{x}_k(t) + D_{u\xi} \B{\xi }(t)+ D_{uy} \B{y}(t) \\
        \B{v}(t) & = C_{v} \B{x}_k(t) + D_{v\xi} \B{\xi}(t) + D_{vy} \B{y}(t)\\
        \B{\xi} & = \phi(\B{v}(t))
    \ea \ee
with $\small{\theta:= \lba{ccc} A_k & B_{\xi} & B_y \\ C_u & D_{u \xi} & D_{uy} \\ C_v & D_{v \xi} & D_{vy} \ear} $ capturing the learnable parameters of $\pi_{\theta}$. In the terminology of \cite{ghaoui2020implicit}, this controller is an ``implicit" recurrent neural network, since $D_{v\xi} \neq 0$ results in a fixed-point equation that implicitly defines the variable $\B{\xi}$. This class of networks encompasses common architectures such as fully connected feedforward neural networks, convolutional layers, and max-pooling layers \cite{ghaoui2020implicit}. By incorporating feedback loops, implicit neural networks are able to achieve the performance of feedforward architectures with fewer parameters \cite{bai2019deep}.

We assume the nonlinearity $\phi$ is applied elementwise so that $\phi(v) = \lba{ccc} \phi_1(v_1) & \cdots & \phi_n(v_n) \ear ^{\top}$ and each $\phi_i$ is sector-bounded. Without loss of generality\footnote{If the sector is given by $[\alpha, \beta] \ne [0,1],$we may apply the loop transformation outlined in \cite[Sec. II-D]{junnarkar2022synthesis} to obtain an equivalent system sector bounded by $[0,1].$}, we take this sector to be $[0,1]$, which is satisfied by common activation functions such as $\mathrm{ReLU}$ and $\tanh$, so that
    \be \label{eq:v_xi_uncertainty}
        \lba{c} v \\ \xi \ear ^{\top} \lba{cc} 0 & -\frac{1}{2}\Lambda \\ -\frac{1}{2}\Lambda & \Lambda \ear \lba{c} v \\ \xi \ear \le 0 
    \ee 
for any diagonal $\Lambda \succ 0$.
The interconnection of plant \eqref{eq:nn_plant} and controller $\pi$ is described by
    \be \ba \label{eq:nn_closed_loop}
        \dot{\B{x}}(t)& = \mathcal{A} \B{x}(t) + \mathcal{B}_w \B{w}(t) + \mathcal{B}_{\xi} \B{\xi}(t) \\
        0 & = \mathcal{C} \B{x}(t) -\B{ v}(t) + \mathcal{D}\B{\xi} (t) \\
        \B{\xi}(t) & = \phi(\B{v}(t)),
    \ea \ee 
where $\B{x}(t) = \small{\lba{c} \B{x}_p(t) \\ \B{x}_k(t) \ear},$ and 
    \begin{equation*} \ba
        & \mathcal{A} =\small{\lba{cc} A_p + B_u D_{uy}C_p & B_u C_u \\ B_yC_p & A_k \ear}, ~\mathcal{B}_{w} =\lba{c}  B_w \\ 0 \ear,\\ &\mathcal{B}_{\xi} = \small{\lba{c} B_u D_{u \xi} \\ B_{\xi} \ear},
         \mathcal{C} = \small{\lba{cc} D_{vy} C_p & C_v \ear},~\mathcal{D} = D_{v \xi}.
    \ea \end{equation*}
{Restrictions are placed on $\mathcal{D}$ to ensure well-posedness of the interconnection, e.g., singular values of $\mathcal{D}$ restricted to be less than one. Nonetheless, there are allowable values of $\mathcal{D}$ that are singular; for these cases, the algebraic constraint can not be inverted to admit a standard ODE form for analysis.}

Applying Proposition~\ref{thm:linear_general}, an $L_2$ gain from $\B{w}$ to $\B{y}$ of $\gamma$ holds for the closed-loop system \eqref{eq:nn_closed_loop} if there exist $P = P^\top \succ 0$, diagonal $\Lambda \succ 0$, and scalar $\lambda \ge 0$ for which the following LMI holds
\setlength{\arraycolsep}{3pt}
\begin{equation*} \ba
\small &\lba{cccc} 
        \mathcal{A}^{\top} P + P \mathcal{A} + \gamma^2 \begin{bmatrix}C_p^\top C_p & 0 \\ 0 & 0\end{bmatrix} & P\mathcal{B}_w & 0 & P \mathcal{B}_\xi  \\
        \mathcal{B}_w^{\top}P & -I & 0 & 0   \\
        0 & 0 & 0 & 0 \\ 
        \mathcal{B}_\xi^{\top}P & 0 & 0 & 0 
    \ear
    \\
    & \preceq \lba{cccc}
        0 & 0 & 0 & 0 \\
        0 & 0 & 0 & 0 \\
        0 & 0 & 0 & -\frac{1}{2} \Lambda \\
        0 & 0 & -\frac{1}{2}\Lambda & \Lambda
    \ear
    + \lambda \begin{bmatrix} \mathcal{C}^\top \\ 0 \\ -I \\ \mathcal{D}^\top \end{bmatrix} \begin{bmatrix} \mathcal{C} & 0 & -I & \mathcal{D}\end{bmatrix}.
\ea \end{equation*}

This LMI is convex in $P, \Lambda,$ and $\lambda$, so that feasibility can be checked numerically given parameters $\theta$. This provides a sufficient condition for a controller $\pi_\theta$ to satisfy an $L_2$ gain bound.

\end{exmp}

\subsubsection{Linear DAE without Uncertainty}
The results presented thus far have provided \emph{sufficient} conditions for dissipativity of uncertain DAE systems, where conservatism is introduced to reduce an associated computational burden. In the linear case with no uncertainty, this conservatism is eliminated, as summarized in the following proposition whose
proof is in Appendix A.
\begin{prop} \label{prop:lin_dissipative}
    The linear DAE system \eqref{eq:lin_uncertain}, with $\B{\xi} = 0$, is dissipative w.r.t. the quadratic supply rate \eqref{eq:quad_s}
    if and only if there exists $\lambda \ge 0$ and a matrix $P \succ 0$  such that
    \setlength{\arraycolsep}{2pt}
        \be \ba \label{eq:lin_proof_lmi}
            & \lba{ccc} A^{\top} P + P A & P B_v & P B_w \\ 
            B_v^{\top} P & 0 & 0 \\
            B_w^{\top} P & 0 & 0 
            \ear \preceq  \\ 
           & \lba{ccc} X_{xx} & 0& X_{xw}  \\0 & 0 & 0 \\
           X_{xw}^{\top} &0 & X_{ww}    \ear{\small{+}} \lambda \lba{c} F^{\top} \\ G_v^{\top} \\ G_w^{\top} \ear \lba{ccc} F & G_v & G_w \ear. 
        \ea \ee 
\end{prop}

\vspace{-2mm}
\section{Case Study: Power Network with Line Failures} \vspace{-2mm}\label{sec:case_study}

We analyze the performance of a wide-area control policy for a power network in the event of a single line failure, referred to as an $N\- 1$ contingency \cite{xue2021dynamic}. Designing a separate control policy for each such contingency one-by-one is computationally burdensome. In fact, for large power networks it would be computationally intractable to perform this computation at the time-scale with which the operating conditions (and thus the dynamics of the system) change. \color{black} Thus, it is advantageous to design 
a single controller and  quickly 
verify its performance for a set of potential contingencies simultaneously. In this section, we use Proposition~\ref{thm:linear_general} for this task applied to the IEEE 39-bus power network \cite{athay1979practical}.

\vspace{-4mm}
\subsection{Power Network \& Line Failure Model}\vspace{-1mm}
The dynamics at each of the $10$ generators in the IEEE 39-bus network are modeled with the classical swing equations \cite{chow2020power}:
    \be \ba \label{eq:nonlin_swing}
        &\dot{\B{\delta}}_i (t) = \Omega \B{\omega}_i(t)  \\
        &\dot{\B{\omega}}_i(t)  = \frac{1}{2H_i} \Big( \B{p}_{mi}(t) ~{ -} D_i \B{\omega}_i(t) - \\
    &~~~~~~~~~~~~~~~~\frac{E_i\B{v}_{Gi}(t) \sin(\B{\delta}_i(t) {\small-} \B{\theta}_{Gi}(t))}{X_{di}} \Big)
    \ea \ee
    where $\tilde{\B{E}}_i(t) = E_i e^{j \B{\delta}_i(t)}$ is the internal voltage, $\tilde{\B{v}}_{Gi} = \B{v}_{Gi}(t) e^{j \B{\theta}_{Gi}(t)}$ is the bus voltage phasor, $\B{p}_{mi}(t)$ is the mechanical input power, and $H_i, X_{di},$ and $D_i$ are the inertia, internal transient reactance and damping constants, all for the $i^{\rm th}$ generator.
Linearization about a power flow solution gives
    \be\ba \label{eq:power_lin_ode}
         \frac{d}{dt} \lba{c} \B{d\delta}(t) \\ \B{d\omega }(t)\ear & = \overline{A} \lba{c} \B{d\delta}(t) \\ \B{d\omega}(t) \ear + \overline{B}_v \B{v}(t) \\
         & ~~~~~+ \lba{c} 0 \\ \B{w}(t) + \B{u}(t) \ear,
    \ea \ee 
where $\B{u}$ and $\B{w}$ represent vectors of control and disturbance signals, respectively, and $\B{d\delta}$ and $\B{d\omega}$ denote vectors of the deviation of $\B{\delta}$ and $\B{\omega}$ from their operating points $\delta_0$ and $\omega_0 = 0$. 
The generator dynamics are coupled through the network via the power flow equations:
{\small{
\setlength{\arraycolsep}{3pt}
    \begin{equation*}\ba
        \left( \lba{cc}Y_{11}  & Y_{12} \\ Y_{21} & Y_{22}  \ear {+} \lba{cc} Y_d & 0 \\ 0 & Y_L \ear \right) \lba{c} \tilde{\B{v}}_G(t) \\ \tilde{\B{v}}_L(t) \ear 
        {=} \lba{c} Y_d \tilde{\B{E}}(t) \\ 0 \ear
    \ea \end{equation*}
}}
where $\tilde{\B{v}}_{Li}(t) = \B{v}_{Li}(t) e^{j \B{\theta}_{Li}(t)}$ is the voltage phasor at load bus $i$, $\small{\lba{cc} Y_{11}  & Y_{12} \\ Y_{21} & Y_{22}  \ear} $ is the network admittance matrix, $Y_d$ is a diagonal matrix whose entries are the inverses of the generator internal transient reactances, $\frac{1}{X_{di}},$ and $Y_L$ is a diagonal matrix whose entries are the constant impedance models of loads in the network. Linearizing this complex-valued equation about the power flow solution and separating real and imaginary components gives the linear, real-valued algebraic constraint
    \be \label{eq:power_lin_alg}
        0  = \overline{F} \lba{c} \B{d\delta}(t) \\ \B{d\omega }(t)\ear + G \B{v}(t),
    \ee 
    where \setlength{\arraycolsep}{3pt}$$\small\B{v}(t) := \lba{cccc} \B{dv}_G(t)^{\top} & \B{d\theta}_G(t)^{\top}& \B{dv}_L(t)^{\top} & \B{d\theta}_L(t)^{\top} \ear^{\top} $$ is the deviation of magnitudes $(\B{v}_G, \B{v}_L)$ and angles $(\B{\theta}_G, \B{\theta}_L)$ of the voltages at generator and load buses from their operating points, respectively. (Appendix C provides expressions for $\bA,\bB_v,\overline{F},$ and ${G}$.) We evaluate performance with the output
    \be \label{eq:power_y}
        \B{y}(t) = \lba{cc} 0 & I \ear \lba{c} \B{d\delta}(t) \\ \B{d\omega }(t)\ear =: \overline{C} \lba{c} \B{d\delta}(t) \\ \B{d\omega }(t)\ear.
    \ee 

{For controller design, we assume access to  \emph{relative} angle measurements, e.g. $d\delta_1 - d\delta_2$ and absolute angular velocity measurements, e.g. $d\omega_1$. A linear, static controller will then take the form \cite{wu2015input}
    \be \label{eq:staticK}
        \B{u}(t)  = K\B{x}(t),
    \ee
where $
        \B{x}(t) = Q \lba{c} \B{d \delta}(t) \\ \B{d \omega}(t) \ear$ is a reduced state vector and
$Q$ is a matrix whose columns form an orthonormal basis for the null space of $\lba{cc} \mathbb{1}^{\top} & 0 \ear$.The corresponding reduced dynamics are 
    \be \ba \label{eq:powerDAE}
       & \dot{\B{x}}(t){\small {=}} A \B{x}(t) + B_v \B{v}(t) + B_w (\B{w}(t) + \B{u}(t))\\
       & 0  = F\B{x}(t) + G\B{v}(t)\\
      &  \B{y}  = C \B{x}(t), 
    \ea \ee 
where $A = Q^{\top} \bA Q ,$ $B_v = Q^{\top} \bB_v,$ $B_w = Q^{\top}\lba{cc} 0 & I \ear^{\top} $, $F = \overline{F} Q$ and $C = \overline{C} Q$. The input-output behavior of \eqref{eq:powerDAE} is equivalent to that of the original system \eqref{eq:power_lin_ode}-\eqref{eq:power_y}, because (i) the DAE \eqref{eq:power_lin_ode}-\eqref{eq:power_lin_alg} is invariant to uniform shifts in angles, i.e., 
 \begin{equation} \label{eq:zero_mode}
        \overline{A} \lba{c} \mathbb{1} \\ 0 \ear = 0,~~\overline{F} \lba{c}\mathbb{1} \\ 0  \ear  + G {\small{\lba{c} 0 \\ \mathbb{1} \\ 0 \\ \mathbb{1} \ear }}= 0,
    \end{equation}
and (ii) the mode corresponding to these uniform angle shifts is unobservable from output \eqref{eq:power_y}.
}

\subsubsection{Line Failure Model \& Controller Design}
\begin{figure}
    \centering
    \includegraphics[width = .5\textwidth]{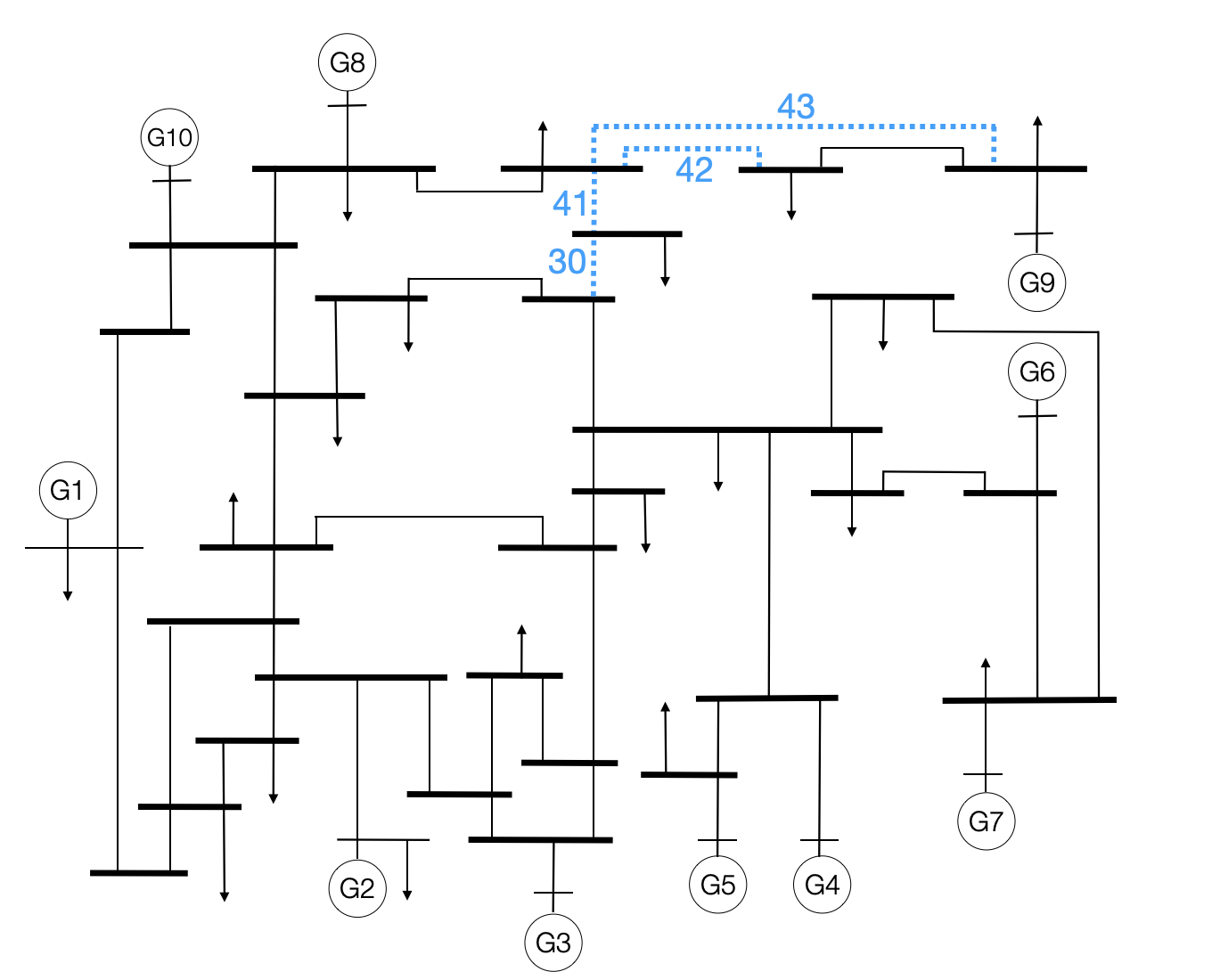}\vspace{-3mm}
    \caption{IEEE 39-Bus Network with potential line failures indicated with dashed blue lines.} \vspace{-5mm}
    \label{fig:ieee39}
\end{figure}
{We restrict our attention to line failures that do not disconnect the underlying graph structure; this corresponds to the vast majority of lines in a typical network \cite{taft2020connectivity}. For instance, we do not consider line failures that result in the disconnection of a generator. In this setting, the dynamics of the perturbed system remain approximately linear.}

We consider four line failures of interest: lines 30, 41, 42, and 43 (see Figure~\ref{fig:ieee39}). We design a controller \eqref{eq:staticK} to place all closed-loop eigenvalues in the half plane $\{z; ~\Re(z) <-0.5 \}$ for the dynamics resulting from line 43 failing. The closed-loop dynamics in this case are approximately 
\begin{subequations}  \label{eq:43withFB} \begin{align} 
        \dot{\B{x}}(t) &= A_{\rm cl} \B{x}(t) + B_v \B{v} (t)+ B_w \B{w}(t)\\
        0 & = F\B{x}(t) + (G + \Delta G_{43}) \B{v}(t) \label{eq:alg431}\\
        \B{y}(t) & = C \B{x} (t),
        \end{align} \end{subequations} 
where $A_{\rm cl}:= (A{\small +}B_u K)$, and the change to the closed-loop dynamics due to the removal of line 43 is approximated by the additive perturbation $ \Delta {G}_{43}$.

{Our objective is to verify $H_{\infty}$ performance of the system in feedback with this wide-area controller for any of the line failures of interest.} Rather than checking each line failure one-by-one, we prioritize computationally tractability 
by constructing a (conservative) uncertainty set containing each contingency, and verifying  dissipativity ($L_2$-gain bound) over this uncertainty set through one LMI using Proposition~\ref{thm:linear_general}.

To create this uncertainty set, we approximate the 
change in the closed-loop dynamics due to the failure of line $i$ as an additive perturbation $\Delta G_i$ to $G$. We compute the differences $\Delta G_{i} \small{-} \Delta G_{43}$ for $i = 30,41,42$. The magnitudes of the singular values of these differences drop off rapidly, and thus we
approximate each  
by a ``structured" \cite{chu2003structured} rank three matrix, where ``structure" corresponds to preserving the physical property \eqref{eq:zero_mode}:
        $ \Delta G_{41}\small{-} \Delta G_{43} = H_{1} J_{1}^{\top}, \Delta G_{42}\small{-} \Delta G_{43} = H_{2} J_{2}^{\top},
         \Delta G_{30}\small{-} \Delta G_{43} = H_{3} J_{3}^{\top},$
with $H_i, J_i \in \mathbb{R}^{78 \times 3}.$
 Then,  
    \be \label{eq:power_alg_uncertain}
        0 = Fx + (G {\small+} \Delta G_{43}) v + \sum_{i=1}^3\tfrac{1-\theta_i }{2} H_i J_i^{\top}v, 
    \ee
$\theta_i \in [-1,1]$, covers the algebraic constraint corresponding to each of the four line failures of interest, e.g., $\theta_1=\theta_2=\theta_3 = 1$ and 
$\theta_1 = \small{-}1, \theta_2 = \theta_3 = 1$ correspond to the removal of line 43 and 41, respectively.
This uncertainty is incorporated as
    \be \ba \label{eq:power_uncertain1}
        \dot{\B{x}}(t) & = A_{\rm cl} \B{x}(t) + B_v \B{v}(t) + B_w \B{w}(t), \\
        0 & = F \B{x}(t) + G_v \B{v}(t) + G_{\xi} \B{\xi}(t), \\
        \B{y}(t) & = C \B{x}(t),
    \ea \ee 
where  
    \be \ba\label{eq:power_uncertain2}
     &G_{\xi} {\small{=}} \lba{ccc} H_1 & H_2 & H_3 \ear,\\ 
       & G_v {\small{=}} G {\small{+}} \Delta G_{43} {\small{+}} \tfrac{1}{2} \left( H_1 J_1^{\top} {\small{+}} H_2 J_2^{\top} {\small{+}} H_3 J_3^{\top} \right)
    \ea \ee 
and $ \B{\xi}(t) = \lba{ccc} \B{\xi}_1(t)^{\top} & \B{\xi}_2(t)^{\top} &\B{\xi}_3(t)^{\top}\ear ^{\top} $ is the output of some $\Delta$ satisfying the pointwise quadratic constraints 
\be \label{eq:power_uncertain3}
        \lba{c} \xi_i \\ v \ear^{\top}\lba{cc} X_i &\frac{-1}{2} Y_i J_i^{\top}\\ \frac{-1}{2}J_i Y_i^{\top} & \frac{-1}{4}J_i X_i J_i^{\top} \ear \lba{c} \xi_i \\ v \ear \le 0 
    \ee 
for any $X_i = X_i^{\top} \succeq 0$ and $Y_i = -Y_i^{\top}$ \cite{megretski1997system}. 

\begin{rem}
    This uncertainty set could be equivalently characterized by a polytopic set, as in \cite[Sec. 3]{uezato1999strict}. Characterization \eqref{eq:power_uncertain3} results in a lower dimensional LMI condition and allows us to incorporate the uncertainty set parameters $X_i, Y_i$ as additional design variables for more flexibility. 
\end{rem}

\vspace{-5mm}
\subsection{$H_{\infty}$ Norm Bound over a set of Line Failures}
We compute a bound on the $L_2$ gain from $\B{w}$ to $\B{y}$ of \eqref{eq:power_uncertain1}-\eqref{eq:power_uncertain2} with 
$\Delta$ satisfying \eqref{eq:power_uncertain3} using Proposition~\ref{thm:linear_general}. We formulate the corresponding convex optimization problem 
\setlength{\arraycolsep}{2.5pt}
    {\small{
    \be \ba \label{eq:power_exmp_opt}
        & \min_{P,\lambda, X_1, Y_1, X_2, Y_2, \gamma^2} ~~~ \gamma^2\\
        &~~~{\rm s.t.}  ~~ P\succ 0, X_i = X_i^{\top} \succeq 0 , Y_i = -Y_i^{\top}, \lambda \ge 0,\\
        &~~~~~~~~~\lba{cccc} A_{\rm{cl}}^{\top} P \small{+} P A_{\rm{cl}} \small{+} C^{\top} C & P B_v &  0 & PB_w\\
        B_v^{\top} P &  0 & 0 &0 \\
         0 & 0 & 0 & 0 \\
        B_w^{\top} P &  0 & 0 & -\gamma^2 I 
        \ear\\
    &~~~~~~~~~~\preceq \lambda \lba{c} F^{\top} \\ G_v^{\top} \\ G_{\xi}^{\top}  \\ 0 \ear \lba{cccc}  F & G_v & G_{\xi} & 0 \ear ~+\\
       &~~~~~~~~~~~\lba{cccc} 
       0 & 0&0&0\\
       0 & \frac{-1}{4} \sum_{i=1}^3 \left( J_{i} X_i J_i^{\top} \right) & W & 0 \\
       0 & W^{\top} &X&0\\
       0 & 0&0&0
       \ear
    \ea \ee}} 
where $W:=\frac{-1}{2} \lba{ccc} J_1 Y_1^{\top} & J_2 Y_2^{\top} & J_3 Y_3^{\top} \ear $ and $X$ is the block diagonal matrix of $\{X_1, X_2, X_3\}.$ We solve \eqref{eq:power_exmp_opt}
numerically in MATLAB using CVX \cite{cvx} with the SeDuMi solver \cite{sturm1999using} resulting in an $L_2$ gain bound of $\gamma = 2.31$ over the uncertainty set. To evaluate conservatism, we compute the $L_2$ gain corresponding to each of the four line removals:
    \begin{center}
    \begin{tabular}{c|c|c|c|c}
        Line removed & 30 & 41 & 42 & 43 \\ \hline
         $\begin{array}{c} \text{Closed-loop} \\ H_{\infty} \text{-norm} \end{array}  $& 2.215 & 2.222 & 2.219 & 2.217\\ 
    \end{tabular}
    \end{center}
We compute $L_2$ gains over the full uncertainty set via a grid search - the maximum value obtained is 2.2719, occurring at $\theta_1 = 0.1, \theta_2 =0, \theta_3 = 0$ in \eqref{eq:power_alg_uncertain}; note that this point does not correspond to a physical line removal.

{We analyze the amount of conservatism introduced by our methodology.}
Our bound $\gamma = 2.31$ 
is 3.98\% over the true maximal $L_2$ gain over the four line removals of interest and 
1.68\% over true bound for the full uncertainty set. 
Thus, for this case study, neither the choice of a larger uncertainty set nor the restriction to 
a single storage function cause 
much conservatism. 
An added benefit of this approach is the implicit incorporation of robustness. 

{We remark on how this methodology compares to an alternate approach of modeling the power network equations as an ODE, rather than a DAE. An advantage of the DAE framework is that it allows for line failures to be modeled as uncertain low rank perturbations to the matrix $G_v$. If we instead utilized the equivalent power network model 
    \be 
        \dot{\B x}(t) = (A_{\rm cl} - B_v G_v^{-1} F){\B x}(t) + B_w {\B w}(t),
    \ee 
perturbations to $G_v$ would need to be inverted to obtain perturbations on the matrix $(A_{\rm cl} - B_v G_v^{-1} F)$, eliminating the structure that allows for simple characterization of the uncertainty set utilized.\\
\indent More generally, whenever the algebraic condition \eqref{eq:nonlin_const} of a DAE is invertible, the dynamics could be equivalently represented by an ODE. This approach will not apply to general DAE systems, whose algebraic conditions may not be invertible. Even with invertibility, the DAE form might be advantageous in (i) preserving structure captured by the algebraic constraint or (ii) avoiding inversion of a poorly conditioned matrix.}

\vspace{-2mm}
\section{Conclusion}
\vspace{-1mm}
A general framework for dissipativity analysis of DAE systems with uncertainties described by IQCs was provided. Numerical methods for verifying sufficient conditions for dissipativity of these systems in the case of polynomial or linear dynamics were illustrated. Analysis of the IEEE 39-bus power system subject to line failures provided a case study that highlighted the amount of conservatism introduced by sufficient conditions could be quite minimal; quantifying or minimizing this conservatism 
could be examined in future work. Further analysis of this case study is also the subject of ongoing work, e.g., efficient methods to group together sets of line failures to characterize by uncertainty sets is currently being investigated \cite{junnarkar2024grouping}.

\bibliographystyle{ieeetr} 
\bibliography{DAE_dissipativity}

\vspace{-1mm}
\section*{Appendix}

\subsection{Proof of Proposition~\ref{prop:lin_dissipative}}\vspace{-1mm}
By the lossless S-procedure (see, e.g., \cite{boyd2004convex}),
the existence of $\lambda \ge 0$ satisfying \eqref{eq:lin_proof_lmi} is equivalent to nonpositivity of 
\be \label{eq:const_quad}
         \lba{c} x \\ v\\ w \ear^{\top}\lba{c} F^{\top} \\ G_v^{\top} \\ G_w^{\top} \ear  \lba{ccc} F & G_v & G_w \ear  \lba{c} x \\ v\\ w \ear 
    \ee 
implying 
{\small{ \be \ba \label{eq:lin_diss_LMI}
      &  \lba{c} x \\ v\\ w \ear^{\top} \Bigg(  \lba{ccc} A^{\top} P + P A & P B_v & P B_w \\ 
            B_v^{\top} P & 0 & 0 \\
            B_w^{\top} P & 0 & 0 
            \ear \\
            & ~~~~~~~~~~~- \lba{ccc} X_{xx} & 0 & X_{xw} \\ 0&0&0\\
           X_{xw}^{\top} & 0 & X_{ww}    \ear\Bigg) \lba{c} x \\ v\\ w \ear \le 0. 
    \ea \ee
    }}
    Since $V(\cdot)$ is 
    continuously differentiable, dissipativity (Definition~\ref{defn:dissipativity}) with $\B{\xi} = 0$ can be confirmed through the 
    differential characterization that:
    \begin{equation*}
        0 = g(x,v,w, 0) ~ \Rightarrow ~ \triangledown V(x)^{\top} f(x,v,w,0) \le 0
    \end{equation*}
for all $x,v,w$. 
Nonpositivity of \eqref{eq:const_quad} is equivalent to $0 = g(x,v,w, 0)$ for linear $g$ of the form \eqref{eq:lin_const}, and condition $\triangledown V(x)^{\top} f(x,v,w,0) \le 0$ reduces to \eqref{eq:lin_diss_LMI} for linear $f$ of form \eqref{eq:lin_ODE} and quadratic $V$.\hfill $\blacksquare$

\vspace{-3mm}
\subsection{Proof of Theorem~\ref{thm:linear_general}:}\vspace{-1mm}
Assume \eqref{linear_thm_eq} holds, and left and right the inequality by $\lba{ccccc} x^{\top}  & v^{\top} & \xi^{\top}& w^{\top} & \psi^{\top} \ear $ and its transpose to arrive at the following condition, which is equivalent to \eqref{eq:nonlinear_general_int}  in this linear setting:
    \begin{equation*}\ba 
       &  x^{\top} P (A x + B_v v + B_w w + B_{\xi} \xi) + (\star)^{\top} P x +\\
       & \psi^{\top} P_{\Delta} \Big( A_{\psi}^{\top} \psi(t) \small{+} B_{\psi} {\small{\lba{c}x(t) \\ v(t) \\ \xi(t)  \ear}}  \Big) + (\star)^{\top} P_{\Delta} \psi(t)\\
       & \le \lba{c} x \\ w \ear^{\top} \lba{cc} X_{xx} & X_{xw} \\ X_{xw}^{\top} & X_{ww} \ear \lba{c} x \\ w \ear \\
       & + \lambda \lba{c} x \\ v  \\ \xi\\ w \ear ^{\top} \lba{c} F^{\top} \\ G_v^{\top}\ \\ G_{\xi}^{\top} \\ G_w^{\top} \ear \lba{cccc} F & G_v  & G_{\xi} & G_w\ear \lba{c} x \\ v  \\ \xi\\ w \ear  \\
       & + \tau \Big( C_{\psi} \psi + D_{\psi} {\small{\lba{c} x \\ v \\ \xi \ear}} \Big)^{\top} M \Big( C_{\psi} \psi + D_{\psi} {\small{\lba{c} x \\ v \\ \xi \ear}} \Big),
    \ea \end{equation*}
where each of term $(\star)$ may be inferred by symmetry. 
The result then follows from Theorem~\ref{thm:nonlin_dissipativity}.
\hfill $\blacksquare$

\vspace{-2mm}
\subsection{Power Network Model Parameters:}\label{app:power}
The ODE parameters are given by
\begin{equation*} \ba 
   & \bA = \lba{cc} 0_{n_g \times n_g} & \Omega \cdot I_{n_g} \\ {\rm diag} (\frac{-E\circ V_G\circ \cos(\delta-\theta_G)}{2 H\circ X_d}) & {\rm diag}(\frac{-D}{2 H})
    \ear \\
    & \bB =  \lba{cccc} 0_{n_g \times n_g} & 0_{n_g \times n_g} & 0_{n_g \times n_l} & 0_{n_g \times n_l} \\ \bB_{21}
     & \bB_{22} & 0_{n_g \times n_l} & 0_{n_g \times n_l}
    \ear,
\ea \end{equation*}
where $ \small {\bB_{21} = {\rm diag}\left( \frac{-E \circ \sin (\delta - \theta_G)}{2 H \circ X_d}\right)}$, and $\small {\bB_{22} = {\rm diag}\left(\frac{E \circ V_G \circ \cos(\delta - \theta_G)}{2 H \circ X_d} \right) }.$
To define $\overline{F}$ and $G$, we decompose the following matrices into real and imaginary parts: \vspace{-3mm}
    \begin{equation*}\ba
        & Y_d = Y^d_{\rm re} + i Y^d_{\rm im},\\
        & Y^G = \lba{c} Y_{11} + Y_d \\ Y_{21} \ear = Y^G_{\rm re} + i Y^G_{\rm im},\\
        & Y^L = \lba{c} Y_{12} \\ Y_{22} + Y_L\ear = Y^L_{\rm re} + i Y^L_{\rm im}.
    \ea \end{equation*}
With this notation, 
\begin{equation*} \ba 
    & \overline{F} = \lba{cc}
        Y^d_{\rm im} \cdot {\rm diag}(E \circ \cos(\delta) & 0_{n_g \times n_g} \\
        0_{n_l \times n_g} & 0_{n_l \times n_g} \\
         Y^d_{\rm im}  \cdot {\rm diag} (E \circ \sin(\delta) & 0_{n_g \times n_g} \\ 
        0_{n_l \times n_g} & 0_{n_l \times n_g} 
    \ear \\
    & {G} = \lba{cccc} G_{V_G} & G_{\theta_G} & G_{V_L} & G_{\theta_L} \ear,
\ea \end{equation*} 
where 
    \begin{equation*}
       {\small{ G_{V_G} = \lba{c}
            Y^G_{\rm re} \cdot {\rm diag} (\cos\theta_G) - Y^G_{\rm im} \cdot {\rm diag} (\sin\theta_G)\\
            Y^G_{\rm re}  \cdot {\rm diag} (\sin\theta_G) + Y^G_{\rm im} \cdot {\rm diag} (\cos\theta_G)
        \ear}}
    \end{equation*}
    \begin{equation*}\ba
       & G_{\theta_G} =\\
       &\small{\lba{c} 
           \sm Y^G_{\rm re} \cdot {\rm diag}\left (V_G \circ \sin\theta_G \right) \sm Y^G_{\rm im} \cdot {\rm diag} (V_G \circ \cos\theta_G)\\
            Y^G_{\rm re} \cdot {\rm diag} \left( V_G \circ \cos\theta_G \right) \sm Y^G_{\rm im} \cdot {\rm diag} \left( V_G \circ \sin\theta_G\right)
        \ear }
    \ea \end{equation*}
    \begin{equation*}
        {\small{G_{V_L} = \lba{c} 
            Y^L_{\rm re} \cdot {\rm diag} \left(\cos\theta_L \right) - Y^L_{\rm im} \cdot {\rm diag} \left( \sin\theta_L \right) \\
            Y^L_{\rm re} \cdot {\rm diag} \left(\sin\theta_L \right) + Y^L_{\rm im} \cdot {\rm diag} \left( \cos\theta_L\right)
        \ear }}
    \end{equation*}
    \begin{equation*}\ba
        &{\small G_{\theta_L} =}\\
        &{\small{\lba{c} 
            \sm Y^L_{\rm re}\cdot {\rm diag} \left( V_L \circ \sin\theta_L\right) \sm Y^L_{\rm im}  \cdot {\rm diag} \left(V_L \circ \cos\theta_L \right) \\
            Y^L_{\rm re} \cdot {\rm diag} \left(V_L \circ \cos\theta_L \right) - Y^L_{\rm im}  \cdot {\rm diag} \left( V_L \circ \sin\theta_L \right)
        \ear }}
   \ea  \end{equation*}

\end{document}